# Robustness Issues of Timing and Synchronization for Free Electron Lasers

Geng Zheqiao*

Independent Researcher, Kohlenweg 40, 5303 Würenlingen, Switzerland
*Corresponding author. *E-mail address:* gengzheqiao@126.com

**Abstract:** Free electron lasers (FEL) require strict time relations for the electron bunch and RF field interaction, which must be precise and deterministic in time. This is guaranteed by the timing and synchronization systems that should be robust under the situations like a power cycle in the master oscillator, timing master, reference frequency distribution devices or low-level radio frequency (LLRF) devices. After the power cycles, the time relations should be kept or be capable to recover quickly to improve the availability of the FEL machine. This article focuses on the robustness of the timing and synchronization systems, such as the time uncertainty of the RF pulse related with the trigger, the phase uncertainty of frequency dividers after power cycles and the race condition between the trigger and clock. The possible solutions to achieve a robust design of the timing and synchronization systems are also discussed.

## 1 Introduction

Free electron laser (FEL) machines, such as the SwissFEL [1] at the Paul Scherrer Institut in Switzerland, the Linac Coherent Light Source (LCLS) [2] at the SLAC National Accelerator Laboratory in USA, and the Spring-8 Angstrom Compact free electron LAser (SACLA) [3] at the Spring-8 in Japan, are based on linear accelerators (Linac) and have a general architecture as Fig. 1.

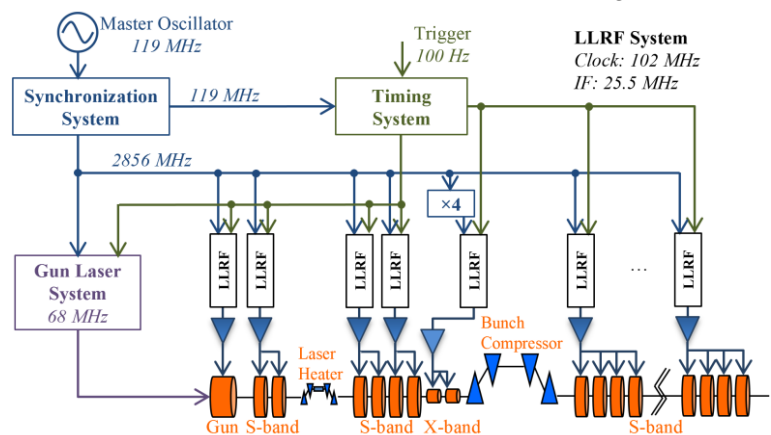

Fig. 1. General architecture of a FEL Linac.

Fig. 1 shows part of the FEL accelerator with a focus on the relations between the master oscillator (MO), timing, synchronization, gun laser and LLRF systems. The timing system [4] generates trigger signals with a picosecond time resolution, and determines the event sequence in a FEL machine to fire different devices at proper time instants. First, the gun laser amplifier is armed and the RF system is fired to build up the RF field in the cavities or structures, then one of the gun laser pulse is picked up and amplified to generate an electron bunch in the RF gun. The synchronization system [5] provides fine time alignments between the gun laser pulse, electron bunch and RF field. When the electron bunch is generated in the RF gun and reaches each downstream cavity or structure, the RF field should have a proper phase to achieve the desired beam energy, bunch length and bunch arrival time. The output of the synchronization system is usually continuous RF reference signals with a femtosecond time resolution. The LLRF system [6-8] is one of the major users of the synchronization signals and maintains the proper RF-beam synchronization. The RF reference signal is used as the original input to the LLRF system, where it is modulated in amplitude and phase, amplified and build up the RF field in the cavities or structures. The LLRF system measures and regulates the RF field. For this purpose, the LLRF system synthesizes other required frequencies like the local oscillator (LO) and clock signals using the RF reference. All frequencies in a FEL machine are derived from a master oscillator to guarantee the synchronization between different devices.

**2 Time Relations in FEL Linac**

2.1 Frequency Assumption for Analysis

To facilitate the time relation robustness analysis, the frequencies of different subsystems are assumed in Fig. 1, which refer mostly the LCLS design [9] and are typical for FELs with S-band normal-conducting Linacs [9]. The master oscillator is assumed to be 119 MHz and the S-band (2856 MHz) and X-band (11.424 GHz) references are generated from it either with frequency multipliers or by selecting proper higher harmonics. The second method is usually used when the MO is a laser oscillator like in the SwissFEL synchronization system [5]. The timing system uses 119 MHz as the event generator (EVG) clock and the trigger rate is assumed to be 100 Hz. The initial trigger input to the EVG is usually generated from the Mains frequency by detecting the zero-crossing points. This allows the entire machine following the same fluctuation of the AC power supply and reduces the relative fluctuation between different pulses. The gun laser oscillator is assumed to generate short laser pulses at 68 MHz repetition rate, from which the 42ed harmonic (2856 MHz) is picked to be phase locked with the reference (see Fig. 3). From the laser oscillator output, the trigger picks up a laser pulse right after its rising edge at 100 Hz, and the picked laser pulse is further processed to generate an electron bunch in the RF gun. In the LLRF system, the clock frequency for the analog-to-digital converter (ADC) and digital-to-analog converter (DAC) is 102 MHz. The intermediate frequency (IF) is assumed to be 25.5 MHz, one-fourth the clock frequency, resulting in an I/Q sampling scheme [10] for the amplitude and phase detection with digital demodulation.

2.2 Time Relations Highlight

The major time relations in a FEL Linac include the trigger synchronization with the master oscillator, the gun laser pulse time w.r.t. the reference and trigger, the LLRF LO and clock phases w.r.t. the reference, the RF pulse starting time w.r.t. the clock and trigger, and the RF phase measurement w.r.t. the LO, clock and trigger.

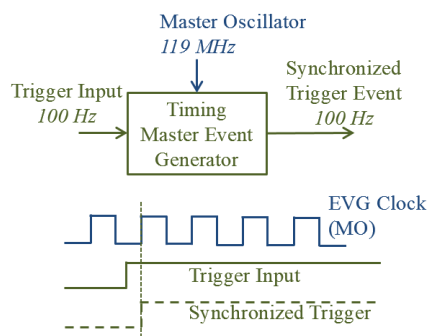

Fig. 2. Event generator trigger synchronization.

The trigger signal derived from the AC Mains is not synchronized with the master oscillator. Before using it as the timing fiducial, the EVG synchronizes the trigger input with the master oscillator signal as a clock (see Fig. 2). This results in a trigger fiducial with its rising edge aligned with one of the rising edges of the clock, and the interval between two triggers becomes an integral multiple of the master oscillator period.

The gun laser is phase locked with the reference signal at one of the harmonics of the laser pulse repetition rate. As shown in Fig. 3, the time interval between two laser pulses at the laser oscillator output is 42 times the reference period. Therefore, the gun laser phase-locked loop (PLL) is equivalent to a frequency divider with a ratio of 1/42. As 100 Hz laser pulses are required for electron bunch generation, the trigger is used to pick up a specific laser pulse that is further amplified, processed and transferred to the RF gun cathode. The trigger delay is adjusted locally to be sure that the trigger locates near the middle between two laser pulses. The actual bunch generation time is defined by both the trigger time and the phase of the 68 MHz signal derived from the laser oscillator output.

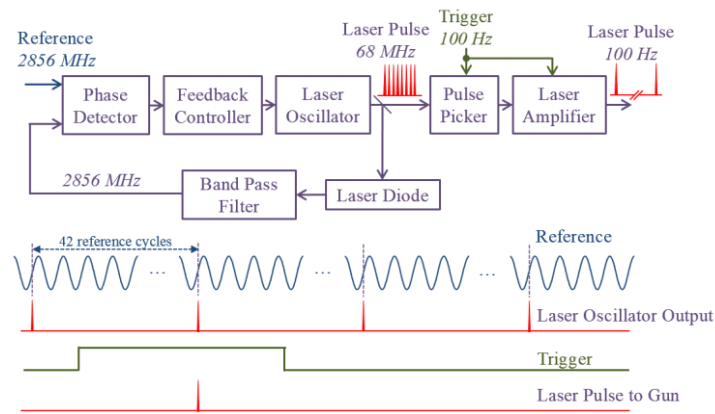

Fig. 3. Gun laser synchronization.

There are different ways to generate LO and clock signals in the LLRF system. Fig. 4 shows one widely used method. Frequency dividers are used to derive the clock and IF signals from the reference input, and the LO signal is generated by mixing the reference and IF signals. Frequency dividers can usually achieve very low noise floor that helps to reduce the added phase noise in the clock and LO signals. The difficult point is the band-pass filter that picks up one of the sidebands in the mixer product. Due to the relatively low IF frequency w.r.t. the reference, a very high-quality-factor filter is required that is usually a cavity filter. A cavity filter can be costly and sensitive to the temperature changes. The time relations between the input and output of the frequency dividers are depicted in Fig. 4. One should be aware that the LO phase follows the IF phase.

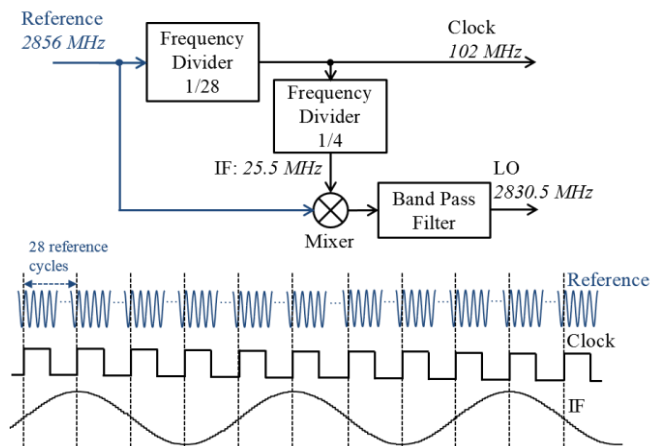

Fig. 4. LLRF LO and clock generation.

For RF stations working in pulsed mode, the RF pulses are started with the trigger distributed by the timing system. Fig. 5 shows a direct up-conversion based RF actuator [11]. The DAC table, which is usually stored in the memory of a field programmable gate array (FPGA), defines the RF pulse envelope. The trigger starts the RF pulse by firing the base band in-phase (I) and quadrature (Q) signals stored in the DAC table via a pair of DACs, with which the amplitude and phase of the reference signal is modulated with a vector modulator. The actual starting time of the RF pulse is aligned with the first clock rising edge (or falling edge) after the trigger time. This implies the dependency of the RF pulse starting time to the trigger time and the clock phase.

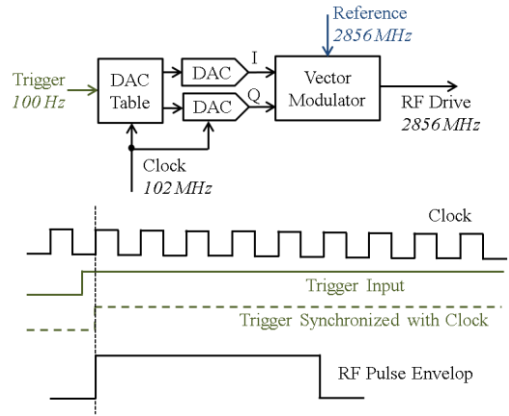

Fig. 5. RF actuation with direct up-conversion.

Fig. 6 shows the time relations in a RF detector based on the I/Q demodulation scheme. The RF signal is down-converted to IF frequency and sampled by the clock. The samples after the trigger can be defined following the sequence "I,Q,-I,-Q,…", where I and Q define the complex envelope of the RF signal, a vector with amplitude and phase. In this scheme, the measured phase is determined by the relative phase between the RF signal, LO, clock and the trigger time.

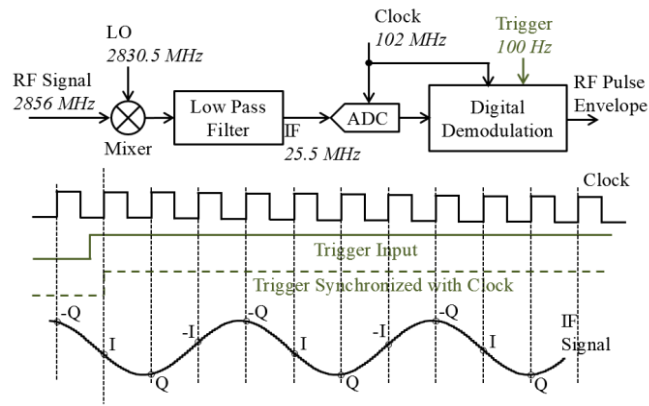

Fig. 6. RF signal detection with I/Q sampling.

Proper acceleration of the electron bunch requires a correct timing between the bunch generation time, arrival time and RF field phase.

2.3 Requirements to Timing and Synchronization

The timing and synchronization systems of existing FELs have put most focuses on the stability [5],[12-13]. The jitter of a trigger should be usually smaller than several picoseconds. The most critical jitter requirement mainly come from the fast kicker magnets which are activated directly by the trigger signal. In some other systems (e.g. LLRF system), the triggers are usually resynchronized with a local low-jitter clock and the jitter requirement is more relaxed. The drift in the timing system is also a concern for some machines and active drift compensation has been implemented in the newest timing system. For many FEL machines, the time resolution of the synchronization should be as low as several femtoseconds. The time precision of the RF-beam synchronization is determined by the desired beam stability (e.g. SwissFEL requires a beam energy stability better than 0.05 %, peak current fluctuation smaller than 5 %, and a bunch arrival time jitter smaller than 20 fs) [14]. The drift in the synchronization system has been specially focused on in the existing designs, in which different active methods for drift compensation are adopted.

Unfortunately, the robustness of the timing and synchronization systems has not been considered much in the existing implementations. Many FEL machines are facing the difficulty to recover the time relations, especially the phase relations between the laser, beam and RF field, after some abnormal situations like a power cycle in different devices. The restart or power cycle of different subsystems is quite usual during maintenance. In order to improve the FEL availability, the time relations should be able to be kept or recovered quickly after the abnormal situations vanish.

In this work, some analysis is performed to identify the abnormal situations and the possible robustness issues of the time relations. Then systematic solutions are proposed to handle these issues, including the frequency selection at the early design stage, the time relation diagnostics and the possible knobs for resynchronization.

## 3 Robustness Issues of Time Relations

3.1 Trigger Period Issue

As shown in Fig. 2, the trigger is synchronized with the MO frequency in EVG, therefore, all the higher harmonics of the MO frequency have integral full cycles between two triggers (see Fig. 7). But for the signals with frequencies lower than the MO frequency, such as the gun laser oscillator output and the LLRF clock and IF signals, the trigger period may not cover full cycles, resulting in different phases of these signals at the trigger times. This situation may cause the following consequence:

1) The waveform pattern of different signals is not stable when viewed in an oscilloscope triggered by the trigger signal.

2) The laser pulse used for electron bunch generation locates at different time after the trigger. This might result in slightly different amplification of the laser pulse if the laser amplifier is started by the trigger and has fixed pattern of ripples in its active time. Due to the relation between the master oscillator frequency (119 MHz) and the laser oscillator frequency (68 MHz), the time difference between the trigger and the laser pulse has 7 discrete values within a laser oscillator period. This is because the trigger meets the same phase of the laser oscillator frequency when it moves by 7 cycles of the master oscillator (7 cycles of 119 MHz covers exactly 4 cycles of 68 MHz).

3) The RF pulse starting time after the trigger has an uncertainty within a DAC clock cycle. If the bunch arrival time is fixed w.r.t. the trigger, the bunch sees different parts of the pulse, resulting in slightly different acceleration if the RF pulse is not perfectly flat. The relation between the master oscillator frequency (119 MHz) and the DAC clock frequency (102 MHz) results in 7 discrete values for the RF pulse starting time.

4) For the RF signal detection, the trigger may select IF samples at different phases as the starting points for demodulation, resulting in a phase uncertainty with integral multiple of 90 degree.

The analysis above indicates that the trigger period should cover full cycles of all the concerned frequencies, which will be realized by pre-synchronizing the trigger with a common subharmonic of all the frequencies (see section 4.2).

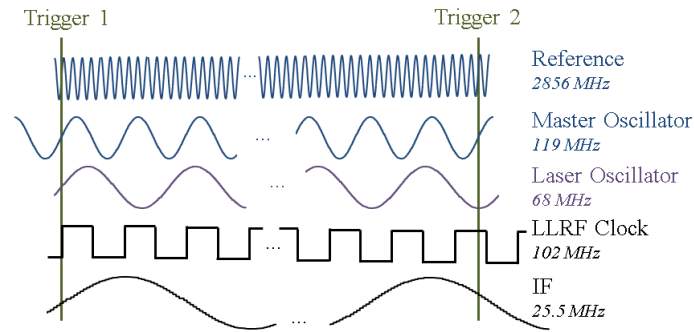

Fig. 7. Pattern of different frequencies between two triggers.

3.2 Frequency Divider Issue

A frequency divider generates an output frequency that is a fraction of the input frequency. Fig. 8 shows the possible output of a 1/3 frequency divider, where the phase has three possible values due to different starting time of the counting.

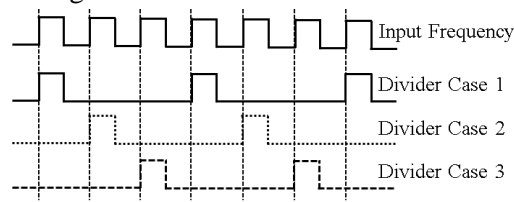

Fig. 8. Phase uncertainty of the output of a frequency divider.

The output phase uncertainty happens when the frequency divider is power cycled or the input signal is interrupted for an uncertain time period. The uncertain phase has $N$ possible values if the frequency division ratio is $1/N$. In the time relations described in section 2.2, the gun laser oscillator and the LLRF LO and clock generator all contain frequency dividers and suffer from the phase uncertainty when they are power cycled or the reference input is interrupted. This issue may cause the following consequence:

1) The laser pulse used for electron bunch generation may be shifted in time by multiple of the reference (2856 MHz) period, which is also denoted as a bucket. If the RF field phase in the gun (also 2856 MHz) is not changed, the electron bunch time shift does not change the accelerating phase but only places the electron bunch at different location of the RF pulse.

   For some FEL machines with a complex frequency strategy, like SwissFEL, the gun laser phase uncertainty may cause more serious problems. The SwissFEL gun laser oscillator is phase locked to 2998.8 MHz which is also the RF frequency in the injector [15]. The phase uncertainty of the gun laser oscillator does not change the accelerating phase in the injector. But in the SwissFEL main Linac, the RF frequency is 5712 MHz, and a bucket change of the gun laser pulse will generate a change of 34.3 degree in the accelerating phase.

2) The LLRF DAC clock has 28 different possible phases (refer to Fig. 4), resulting in different RF pulse starting time within a clock period if the trigger time is fixed.

3) In Fig. 4, the relative phase between the ADC clock and the IF signal used for LO generation has four different possible values, resulting in a phase detection uncertainty with integral multiple of 90 degree.

Power cycle of a frequency divider is a major source of the time relation uncertainty. It can be either reset with a common timing reference or re-synchronized with a phase shifter. The solutions will be discussed in section 4.

3.3 Race Condition

Race condition is a special case for the relative timing between two signals (e.g. trigger and clock). When the rising edges (assume the rising edge is used for timing) of the two signals are aligned close to each other, the sequence between them may toggle due to their time jitter, resulting in jumps in the time relation.

The race condition in the gun laser synchronization (see Fig. 3) selects one of the two adjacent laser pulses, resulting in a bunch arrival time jump by a period of the laser oscillator frequency. As the RF frequency in the downstream RF stations is a harmonic of the laser oscillator frequency, the beam accelerating phase will not change, but the bunch will meet different part of the RF pulse, which introduces acceleration errors if the RF pulse is not fully flat.

In the RF actuator in Fig. 5, the race condition between the trigger and clock generates a jump in the RF pulse starting time by a clock period. For the RF detector in Fig. 6, a jump of the clock synchronized trigger selects a different sample as the starting point of demodulation, resulting in a phase jump of ±90 degree.

**4 Strategies for Robust Timing and Synchronization**

Robust time relations in a FEL Linac against power cycles and signal interruptions are important for the availability of the machine. It is expected that the beam operation can be quickly restored after a maintenance or trouble shooting of the devices in the timing, synchronization, gun laser and LLRF systems. The uncertain time relations should be able to be restored, or at least diagnosed, so that the machine recovery time can be minimized. In this section, a systematic strategy will be introduced to achieve a robust timing and synchronization in the FEL Linac.

4.1 Frequency Selection

Generally speaking, the more types of frequency are used in the machine, the more difficult to achieve robust time relations. The frequencies in different subsystems should be selected systematically to simplify the frequency relations. Minimizing the frequency types is a good guideline to start the frequency selection. With this consideration, the frequencies in Fig. 1 can be simplified to change both the laser oscillator frequency and the LLRF clock frequency to 119 MHz. This can at least avoid the uncertainty in the laser pulse time w.r.t. the trigger and the bunch arrival time w.r.t. the RF pulse starting time if the trigger time is changed by integral period of the EVG clock. Of course, with the new ADC clock frequency, the RF detection should use the non-I/Q demodulation algorithm [10] because the sampling frequency is no longer four times the IF frequency.

Practically, the frequency selection is a trade-off between the robust time relation and the required RF detection bandwidth (determines the minimum IF frequency), the phase noise (determines the minimum MO frequency), the technical difficulty and the cost (e.g. the LLRF clock frequency is limited by the cost and availability of the FPGA, ADC and DAC).

4.2 Trigger Synchronization with Common Subharmonic

To handle the trigger period issues described in section 3.1, the trigger input to the EVG can be pre-synchronized with a common subharmonic of all the concerned frequencies. A period of the common subharmonic covers integral multiples of full cycles of all the involved frequencies. For Fig. 1, the largest common subharmonic is 8.5 MHz, and the EVG trigger synchronization in Fig. 2 changes to Fig. 9 below. With this new synchronization scheme, the trigger period becomes integral multiple of the common subharmonic period, and all the frequencies in Fig. 7 will have the same phases at each trigger, as depicted graphically in Fig. 10.

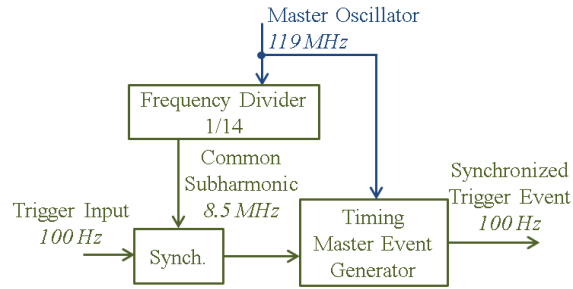

Fig. 9. EVG trigger synchronization with the common subharmonic.

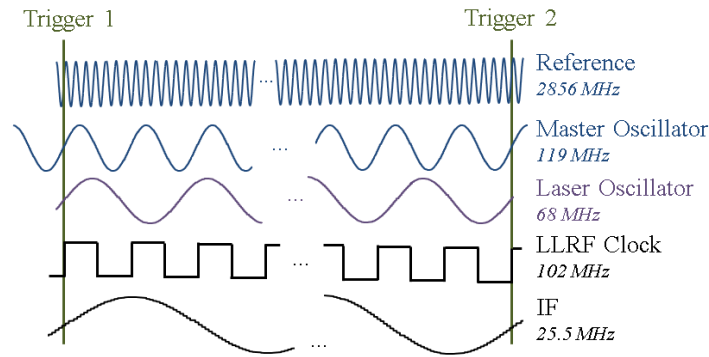

Fig. 10. Pattern of different frequencies between two common subharmonic synchronized triggers.

Due to the phase fluctuation of the AC Mains, the trigger input in Fig. 9 has a small period fluctuation that is discretized by the common subharmonic. Its frequency should not be too small so that the synchronized trigger follows the trigger input with a fine time resolution. The simpler the frequency relations, a higher common subharmonic frequency is applicable.

Please note that we have introduced a frequency divider to generate the common subharmonic from the master oscillator signal. The power cycle of this frequency divider also generates phase uncertainty in its output, resulting in the trigger time to be shifted by integral multiple of the 119 MHz period compared to the expected trigger time. In this case, the problems described in section 3.1 still happen once right after the frequency divider cycling and appear as static variations for the beam operation: a variation of the laser pulse time w.r.t. to the trigger, a variation of the bunch arrival time w.r.t. the RF pulse starting time and a constant phase jump in the RF signal detection. Of course, if the trigger time happens to be very close to the laser pulse or the LLRF clock rising edge, the race conditions can also happen. These static variations can only be mitigated in some special cases (see section 4.3 and 4.4), otherwise, they will appear as residual static errors that should be compensated by other methods like beam based feedback.

4.3 Frequency Divider Resetting

A resettable frequency divider is helpful to correct the phase uncertainty after a power cycle or an interruption of the input signal. The reset signal restarts the counter from zero and reinitializes the output phase. Therefore, if all the frequency dividers are reset with a common signal, the relative phase of their output can be maintained even they are power cycled.

The trigger synchronized with the common subharmonic can be used to reset the frequency dividers. Here we assume the frequency dividers in the LO and clock generator as well as the laser oscillator are all resettable by the trigger signal. As the trigger period covers full cycles of all the frequencies, the pattern of different frequencies will keep the same as in Fig. 10 when the frequency dividers are reset. Therefore, the laser pulse time w.r.t. to the trigger and the bunch arrival time w.r.t. the RF pulse starting time will be all kept. And because the relative phase between the ADC clock and the IF signal used for LO generation are reserved, the phase measurement will be also totally restored by resetting the frequency dividers. In addition, the relative phase between the laser oscillator output (68 MHz) and the RF frequency (2856 MHz) is also restored, resulting in no changes in the accelerating phase.

A special frequency divider is the one used to generate the common subharmonic in Fig. 9. There are no signals to reset it because it provides the very original timing reference. As mentioned before, when this frequency divider is power cycled, the trigger time will have a jump by integral multiple of the 119 MHz period, but this will not affect the beam operation if the LLRF frequency dividers and the laser oscillator are reset by the trigger signal.

The resettable frequency dividers are very useful for robust time relations. But practically, resettable frequency dividers for frequencies higher than hundreds MHz are not much available in the market. In addition, resetting the laser oscillator at each trigger might not be practical. These all require developing other methods to mitigate the phase uncertainty of frequency dividers.

4.4 Robust Phase Detection

One of the major tasks of the LLRF system is to maintain the RF field phase w.r.t. the RF reference phase. This is usually achieved with a phase feedback loop. As mentioned before, the exceptions in the time relations between the trigger, LO and clock generate phase measurement errors which are transmitted to the RF field if the phase feedback is closed. Fortunately, such phase measurement uncertainty can be mitigated with the so-called "reference tracking", see Fig. 11.

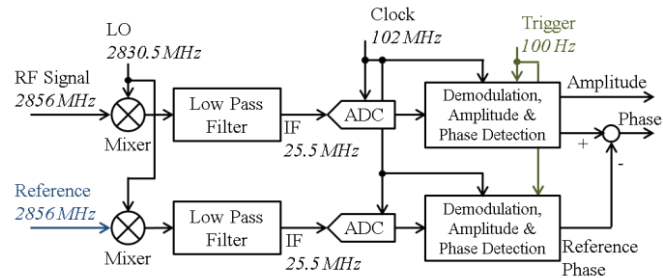

Fig. 11. RF signal phase measurement with reference tracking.

If the RF reference signal is measured with the same LO, clock and trigger signals as the RF signal, both will experience the same phase jump caused by the time uncertainty. By subtracting the phase of the reference signal, such common-mode phase error can be removed. Of course, the reference phase should be low-pass filtered so that not too much high frequency phase noise of the reference signal is passed to the RF signal measurement. The article [16] has more detailed discussion on the topic of robust RF detection.

4.5 Time Relation Diagnostics

Even if the time relations cannot be fully restored in some situations, like after power cycling of the master oscillator, comprehensive diagnostics are helpful to tell which time relations are changed. This helps to improve the observability of the machine.

Fig. 12 shows the possible diagnostics of the concerned time relations. Here we assume the master oscillator frequency is distributed by the timing system as EVG clock and reconstructed by the event receivers (EVR). The reconstructed EVG clock has always consistent phase with the master oscillator and can be measured with the LLRF system at each RF station. At the RF gun station, the fundamental frequency of the gun laser oscillator can be also measured to detect the relative phase between the master oscillator and the laser oscillator.

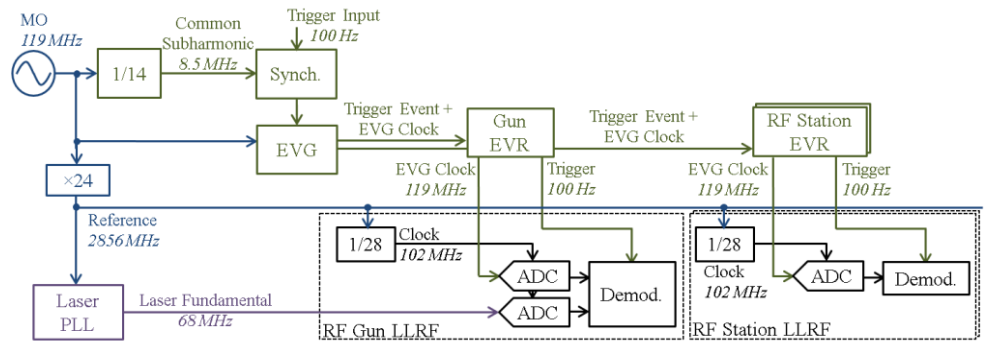

Fig. 12. Diagnostics of the time relations.

In the diagnostics plan in Fig. 12, the EVG clock (119 MHz) and the laser oscillator signal (68 MHz) are directly sampled by the LLRF ADCs clocked at 102 MHz, see Fig. 13. The digital demodulation can follow the non-I/Q algorithm [16]. For the measurement of the EVG clock phase, 6 ADC samples cover 7 periods of the EVG clock, and a shift of an ADC sample results in a phase jump of 420 degree (or 60 degree if unwrapped). For the laser phase measurement, 3 ADC samples cover 2 periods of the laser oscillator fundamental frequency, and each sample corresponds to a phase step of 240 degree. The demodulation is started by the trigger signal. These parameters help to formulate the non-I/Q demodulation configuration.

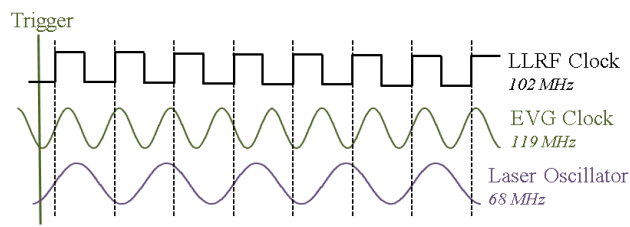

Fig. 13. Direct sampling of the EVG clock and the laser oscillator signal.

The relative timing between the master oscillator and gun laser is only determined by the equivalent frequency division in the laser PLL. This is because the laser PLL reference is derived from the master oscillator and is always phase locked. But the measured phases of the EVG clock and laser oscillator signal are determined not only by the relative timing between them, but also by the ADC clock phase and trigger. By observing the waveforms in Fig. 13, it can be seen that either the LLRF clock phase change or the trigger time change moves the demodulation starting samples on the EVG clock and laser oscillator signal by the same time. This results in phase measurement changes following the frequency relation of the two signals: $119/68 = 7/4$. On the other hand, the laser phase uncertainty due to the laser PLL frequency division is multiple of 8.57 degree at 68 MHz. These facts can be used to judge the phase measurements of the EVG clock and laser oscillator signal. When the Linac is in normal operation, the phases of the two signals can be recorded as a reference. The phase measurement changes, which happen usually after an exception in the Linac, should follow the ratio 7/4, otherwise, the laser PLL has jumped by a number of 2856 MHz buckets that can be estimated with the phase relations given above.

In each RF station, the EVG clock phase measurement offers information about the time changes in the trigger and ADC clock. As shown in Fig. 12, the EVG clock and the reference signal input to the ADC clock generator have always consistent phase w.r.t. the master oscillator. The EVG clock phase measurement reflects the uncertainty in the trigger time and the frequency divider in the clock generator. If the frequency divider is power cycled, the 102 MHz clock shifts by integral periods of 2856 MHz, resulting in a phase measurement jump by integral of 15 degree for 119 MHz. The trigger time jumps by integral periods of master oscillator when the frequency divider for common subharmonic is power cycled. It may select different ADC samples to start demodulation and generate phase measurement jumps by multiple of 420 degree. Some meaningful information can be better achieved by checking the EVG clock phase measurement from all RF stations:

1) If the measured EVG clock phase at only one RF station changes, it should be caused by the ADC clock phase or trigger time change at this RF station. The source can be identified from the phase change value.

2) If the measured EVG clock phases at almost all RF stations change, the source may be the common subharmonic frequency divider if the phase jumps are integral of 60 degree, or the master oscillator (e.g. power cycles) if the phase jumps have more complicated values.

4.6 Time Relation Resynchronization

With the time relation diagnostics, the changes in some important time relations can be corrected by inserting proper knobs, such as a delay stage or a phase shifter. Fig. 14 shows the strategy to resynchronize the laser PLL with a phase shifter in the reference input path. When the phase shifter is rotated by a full circle, the output of the laser oscillator will be moved by one bucket of the reference signal. This method can be used to restore the relative phase between the master oscillator and the laser oscillator. One should be aware that the moving speed of the reference phase shouldn't be so large that the laser oscillator loses locking. The same method can be used to resynchronize other frequency dividers if necessary.

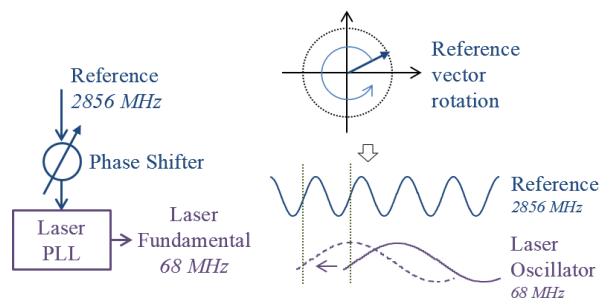

Fig. 14. Resynchronization of the laser phase. The waveforms are for demonstration and do not reflect the real period of the signals.

4.7 Race Condition Detection and Handling

The race condition between trigger and clock introduces random uncertainty during normal operation. In some sense, this is more harmful than frequency dividers because their time uncertainty happens only once after power cycles. In a FPGA, the race condition between trigger and clock can be detected with three flip-flops as depicted in Fig. 15. In this method, the trigger is sampled by both the rising and falling edges of the clock. If there is no race condition, the trigger appears in a fixed sequence at the outputs of the two flip-flops (Q1 and Q2), and the race condition flag output from the third flip-flop has always the same value (0 or 1). If the race condition happens, the sequence of the rising edges of Q1 and Q2 will toggle, so as the value of the race condition flag. The race condition flag can be checked after each trigger, and the race condition can be detected if its value toggles.

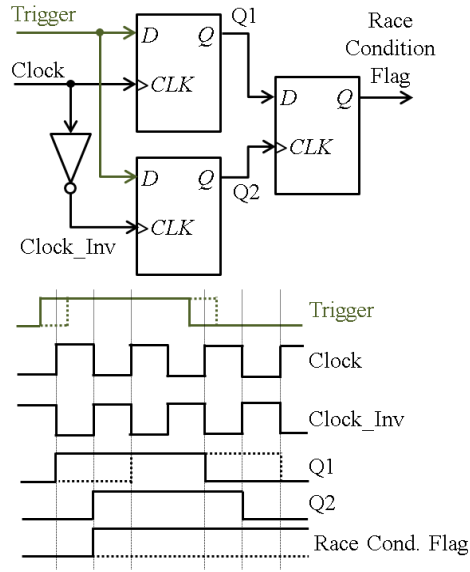

Fig. 15. Detection of the race condition between trigger and clock.

To handle the race condition, one can insert a piece of cable in the trigger or clock path. In FPGA, the race condition can be handled by synchronizing the trigger with the clock falling edge instead of the rising edge. As a consequence, the trigger time will be moved by half of the clock period.

4.8 Residual Time Error Handling

For the Linac in Fig. 1, the following things should be at least implemented to guarantee a successful beam acceleration after an exception in time relations (e.g. master oscillator power cycle):
1) Synchronize the trigger with common subharmonic.
2) Implement reference tracking for RF phase detection.
3) Diagnose and resynchronize the laser phase w.r.t. the master oscillator.
4) Detect and handle race conditions.

With the minimum implementation above, the issues like the uncertainty in laser pulse time w.r.t. the trigger and the bunch arrival time w.r.t. the RF pulse are still not resolved. These issues will generate errors in the RF field felt by the beam that are usually small and do not stop the beam transmission. These small errors become static after the time relations get stable again, and can be compensated by the beam based feedback [17]. One should be aware that these residual errors make the Linac very difficult to completely restore the beam acceleration even the drift in the machine is perfectly compensated.

Furthermore, synchronizing the MO with an external uninterruptible signal source (e.g. GPS) is helpful to guarantee the MO continuity after power cycles. For critical devices like the common subharmonic frequency divider, an uninterruptible power supply (UPS) is a good choice to avoid abnormal power drops.

**5 Conclusions and Outlook**

This article provides a systematic analysis of the time relations, the robustness issues and the possible solutions for a FEL Linac. Some of the solutions have been implemented in LCLS and SwissFEL. Achieving robust timing and synchronization requires adopting the solutions systematically at the design stage of a FEL machine. It will be difficult to change frequencies, add extra time relation diagnostics and place adjustment devices after the machine construction is done. The analysis and solutions provided in this article can also be referred in the design of the timing and synchronization systems of other kinds of accelerators (e.g. circular accelerator). The robustness of the time relations should be emphasized if one expect the beam to be restored quickly after a power cycle of the timing, synchronization and LLRF devices.